\documentclass[aps,pra,twocolumn,longbibliography]{revtex4-1}
\usepackage{amsmath}
\usepackage{amssymb}
\usepackage{comment}

\newcommand\beq{\begin{equation}}
\newcommand\eeq{\end{equation}}
\newcommand\bea{\begin{eqnarray}}
\newcommand\eea{\end{eqnarray}}
\newcommand\nn{\nonumber}

\newcommand{\bra}[1]{\langle #1 |} 
\newcommand{\ket}[1]{| #1 \rangle }

\newcommand\ro{\hat\rho}
\newcommand\ox{\otimes}
\newcommand\Ho{\hat H}
\newcommand\VoG{\hat{V}^\mathrm{G}}

\newcommand\tr{\mathsf{tr}}

\newcommand{\Schr}{Schr\"odinger}
\newcommand\Nlim{N\rightarrow\infty}

\newcommand{\Si}{\Sigma}
\newcommand\Sii{_{\Si_0}}
\newcommand{\Dc}{\mathcal{D}}
\newcommand\muo{\hat{\mu}}
\newcommand\To{\hat{T}}
\newcommand\Psib{\bold{\Psi}}
\newcommand\rv{\mathbf{r}}
\newcommand\sv{\mathbf{s}}

\begin{document}
\title{Semiclassical world is one of infinite many cloneworlds in common spacetime}
\author{Lajos Di\'osi}
%\email{diosi.lajos@wigner.hu}
%\homepage{www.wigner.hu/~diosi} 
\affiliation{Wigner Research Center for Physics, H-1525 Budapest 114 , P.O.Box 49, Hungary}
\affiliation{E\"otv\"os Lor\'and University, H-1117 Budapest, P\'azm\'any P\'eter stny. 1/A, Hungary}
\date{\today}

\begin{abstract}
We consider $N$ clones of the quantized world, interacting with each other via quantum gravity, 
coupled by the downscaled Newton constant $G/N$.  In the limit
 $\Nlim$, we obtain the semiclassical Einstein equation for every single cloneworld.
In the non-relativistic limit, De Filippo had already obtained the semiclassical \Schr--Newton equation,
we present an alternative elementary proof.  In the general relativistic case we complete the semi-finished
derivation of Hartle and Horowitz. We compare our simple correlated cloneworlds with Stamp's more
complicated proposal of correlated worldlines and show why the two constructions differ despite the 
conceptual similarity.
\end{abstract}

\maketitle

\section{Introduction}
Although quantum theory was initially conceived for the atomic world, it has come to be thought of as 
the universal theory of the whole Universe. Since the quantization of Einstein's gravity theory has not
yet been solved, we are forced to use the semiclassical theory where gravity remains classical 
(unquantized) and it interacts with the quantized matter. The simplest semiclassical theory goes
back to the 1960s  \cite{moller1962,rosenfeld1963}.

Consider a given foliation of the spacetime in spacelike hypersurfaces $\Si$.
The statevector of the quantized matter evolves with the Tomonaga--Schwinger equation \eqref{TSE}
where the Hamiltonian density depends on the classical metric $g_{ab}$ which is the solution of
the semiclassical Einstein equation \eqref{sclEE} with the Einstein tensor
on the left and the expectation value of the energy-momentum operator on the right: 
\bea
\label{TSE}
\frac{\delta\ket{\Psi_\Si}}{\delta\Si(x)}&=&-i\hat{\mathcal{H}}(x)\ket{\Psi_\Si},\\
\label{sclEE}
G_{ab}(x)&=&8\pi G\bra{\Psi_\Si}\To_{ab}(x)\ket{\Psi_\Si},~~~~~~(x\in\Si).
\eea
In the Newtonian limit the semiclassical theory becomes much simpler. The statevector of the quantized
non-relativistic matter evolves with the \Schr~ equation \eqref{SchE} where $\Ho$ is the self-Hamiltonian
and the non-relativistic mass distribution operator $\muo$ couples to the classical Newton potential $\Phi$
which is the solution of the Poisson--Newton equation \eqref{Poiss}: 
\bea
\label{SchE}
\frac{d\ket{\Psi_t}}{dt}&=&-i\left(\Ho+\int\Phi(\rv,t)\muo(\rv)d\rv\right)\ket{\Psi_t},\\
\label{Poiss}
\Delta\Phi(\rv,t)&=&-4\pi G\bra{\Psi_t}\muo(\rv,t)\ket{\Psi_t}.
\eea
This equation, unlike its general relativistic form \eqref{sclEE}, is easy to solve. Let us insert the solution into eq. \eqref{SchE}: 
\beq\label{SNE}
\frac{d\ket{\Psi}}{dt}=-i\Ho\ket{\Psi}+iG\int\int\muo(\rv)\bra{\Psi}\muo(\sv)\ket{\Psi}\frac{d\rv d\sv}{\vert\rv-\sv\vert}\ket{\Psi}.
\eeq
Although this equation had already been used  for quantized stellar masses \cite{ruffini1969systems},
its possible relevance in foundations and its features in the quantized motion of nanomasses 
were revealed by the present author in 1984 and by Penrose who called it  
the \Schr--Newton equation \cite{diosi1984,penrose1996}.

In 1981, Hartle and Horowitz considered $N$ identical bosonic fields coupled by quantized gravity at downscaled 
coupling $G/N$ \cite{hartle1981ground}. 
The leading order in $1/N$ yielded an approximation closely related to semiclassical gravity:
\beq
\label{sclHH}
G_{ab}(x)=8\pi G\frac{\bra{0_+}\To_{ab}^H(x)\ket{0_-}}{\bra{0_+}0_-\rangle}
\eeq 
where $\ket{0_\pm}$ are the asymptotic initial/final bosonic vacuum  states, respectively, $\To_{ab}^H(x)$
is in the Heisenberg  picture.  This differs from the correct semiclassical equation 
$G_{ab}(x)=8\pi G\bra{0_-}\To_{ab}^H(x)\ket{0_-}$, as noticed by the authors.

Twenty years later and being apparently unaware  of the construction \cite{hartle1981ground}, 
De Filippo discussed its non-relativistic special case \cite{de2001schroedinger}. 
Let our system of interest exist in $N$ identical copies. Let them interact via the Newton pair potential 
with the downscaled Newton constant $G/N$. 
Consider the \Schr~equation and take an uncorrelated initial state $\ket{\Psi}^{\ox N}$: 
\bea
\label{dPsi}
\frac{d\ket{\Psi}^{\ox N}}{dt}&=&-i\left(\sum_n\Ho_n+\frac{1}{N}\sum_{n<m}\VoG_{nm}\right)\ket{\Psi}^{\ox N},\\
\label{Vnm}
\VoG_{nm}&=&-G\int\int\frac{\muo_n(\rv)\muo_m(\sv)}{\vert\rv-\sv\vert}d\rv d\sv.
\eea
$\Ho_n$ stands for the same Hamiltonian $\Ho$ acting on the $n^{th}$ subsystem, and similarly $\muo_n(\rv)$ is the mass distribution operator of the  $n^{th}$ subsystem. 
The \emph{reduced dynamics} of any one of the  $N$ components is the same, let us take the $1^{st}$ one:
\beq\label{dPsiPsi}
\frac{d}{dt}\left(\ket{\Psi}\bra{\Psi}\right)=\left(\prod_{n\neq1}\tr_n\right)\frac{d}{dt}\left(\ket{\Psi}\bra{\Psi}\right)^{\ox N}.
\eeq
In a lenghty path-integral proof , 
De Filippo showed that in the $\Nlim$ limit this reduced state remains a pure state and its evolution is governed
by the \Schr--Newton equation \eqref{SNE}. 

Here we present a much simpler proof. It starts with the reduced dynamics 
of a fixed number $k<N$ of copies: 
\beq\label{dPsiPsik}
\frac{d}{dt}\left(\ket{\Psi}\bra{\Psi}\right)^{\ox k}=\left(\prod_{n>k}\tr_n\right)\frac{d}{dt}\left(\ket{\Psi}\bra{\Psi}\right)^{\ox N}.
\eeq
Using the eqs. \eqref{dPsi} and \eqref{Vnm} together with \eqref{Poiss}, the r.h.s. reads:
\bea\label{dPsiPsiksep}
%\frac{d}{dt}\!\left(\ket{\Psi}\bra{\Psi}\right)^{\ox k}
&&-i\sum_{n=1}^k\left[\Ho_n+\frac{N-k}{N}\int\Phi(\rv,t)\muo_n(\rv)d\rv,\left(\ket{\Psi}\bra{\Psi}\right)^{\ox k}\right]\nn\\
&& -\frac{i}{N}\sum_{n,m=1}\left[\VoG_{nm},\left(\ket{\Psi}\bra{\Psi}\right)^{\ox k}\right].
\eea
In the limit $\Nlim$, the eq. \eqref{dPsiPsik} reduces to this:
\bea
&&\frac{d}{dt}\left(\ket{\Psi}\bra{\Psi}\right)^{\ox k}=\nn\\
&=&-i\sum_{n=1}^k\left[\Ho_n+\int\Phi(\rv,t)\muo_n(\rv)d\rv,\left(\ket{\Psi}\bra{\Psi}\right)^{\ox k}\right]\!\!.
\eea
We have thus proved that a constant number of copies  
will evolve separately by the \Schr--Newton equation \eqref{SchE} each. 
Note that the whole composite of $N$ copies becomes entangled by the \Schr~ equation \eqref{dPsi},
only the constant-size subsystems remain disentangled in the limit $\Nlim$. This explains the general asymptotic
mechanism of semiclassicality's emergence from unitary dynamics.

Section \ref{II} contains our main result. We complete the semi-finished proof of Hartle and Horowitz \cite{hartle1981ground}
and suggest the narrative of correlated clonewords (CCW) in footprints of refs.  \cite{hartle1981ground,de2001schroedinger}.
More recently, also Stamp proposed infinite many clones of physical fields coupled by Einstein gravity \cite{stamp2015rationale}.
Our work enjoys strong motivations by his correlated worldlines  (CWL) theory. 
Section \ref{III} compares it briefly with our CCW theory.

\section{Correlated cloneworlds}\label{II}
The following narrative can be imagined behind the model.  
Suppose that in the \emph{same} quantized spacetime there exist infnite many \emph{identical} quantized worlds
(\emph{cloneworlds}) and we live in one of them. Which one, does not matter, they are all identical. 
Of course, we must replace Newton's coupling $G$ between matter and spacetime curvature  by $G/N$ 
while the number $N$ of cloneworlds is going to the infinity. 

Exact methods are hopeless because quantization of gravity is not yet solved.
We restrict ourselves for the naive form of Feynman's path integrals and
disregard the unsolved problems like the non-renormalizability and   
we disregard even solved ones like the diffeomorphism ambiguity of the metric $g$. 
 
For simplicity, we consider bosonic matter fields $\phi(x)$ only. We begin with $N$ cloneworlds.
Let a spacelike foliation of the spacetime be given and let  the (bosonic) matter in each cloneworld
have the same initial wavefunction(al) $\psi\Sii[\phi\Sii]$ on a hypersurface $\Si_0$. 
Then the joint initial state of the $N$ cloneworlds and their common spacetime reads:
\beq\label{Psii}  
\Psib\Sii[\phi\Sii^1,\dots,\phi\Sii^N,g\Sii]=\left(\prod_{n=1}^N \Psi\Sii[\phi\Sii^n]\right)\Psi\Sii^G[g\Sii],
\eeq
where $\Psi\Sii^G[g\Sii]$ is the initial wavefuncion of the spacetime metric $g$.
The following naive Feynman integral expresses the state on a later hypersurfaces $\Si$:
\bea\label{Psif}
\Psib_\Si[\phi_\Si^1,\dots,\phi_\Si^N,g_\Si]&=&\int\exp\Bigl(iNS_G[g]+i\sum_n S_M[\phi^n,g]\Bigr)\nn\\
&&\!\!\!\!\!\!\!\!\!\!\!\!\!\!\!\!\times\left(\prod_n \Psi\Sii[\phi\Sii^n]\Dc\phi^n\right)\Psi\Sii^G[g\Sii]\Dc g.
\eea
Here $S_G$ is the Einstein--Hilbert action, $S_M$ is the action of the matter in each cloneworld.
Consider  the following standard Feynman integral:
\beq\label{Psifsingle}
\Psi_\Si[\phi_\Si;g]=\int\exp\left(iS_M[\phi,g]\right)\Psi\Sii[\phi\Sii]\Dc\phi.
\eeq
This expresses  the unitary evolution of the matter's wavefunction in one  world 
in the  background metric $g$. It is known that $\Psi_\Si$ satisfies the
Tomonaga--Schwinger equation \eqref{TSE} where $\hat{\mathcal{H}}$ depends on $g$.
We are going to show in the limit $\Nlim$ that the state
of quantized matter in each cloneworld keeps to be the pure state  $\Psi_\Si[\phi_\Si;g]$
where the metric $g$ depends on these pure states via the semiclassical Einstein eqauation \eqref{sclEE}.
Just to be clear: the evolution  \eqref{Psifsingle} is unitary as long as $g$ is an
independently fixed geometry.  The evolution is not unitary and not even linear in the semiclassical
model. 

Recognizing the structures \eqref{Psifsingle} in the expression \eqref{Psif} of 
the total state $\Psib_\Si$, we can rewrite the r.h.s. of \eqref{Psif}:
\bea\label{Psifshort}
&&\Psib_\Si[\phi_\Si^1,\dots,\Phi_\Si^N,g_\Si]=\\
&=&\int\exp\left(iNS_G[g]\right)\left(\prod_n \Psi_\Si[\phi_\Si^n;g]\right)\Psi\Sii^G[g\Sii]\Dc g\nn.
\eea
At this very stage we modify the  method of effective action used by Hartle and Horowitz \cite{hartle1981ground}.
We do it in  such way that fits to calculating reduced dynamics of a single cloneworld, let it be the $1^{st}$ one. 
Its reduced density matrix is defined by
\bea\label{reddyn}
\rho_\Si[\phi_\Si^1,\phi_\Si'^1]
&=&\int\!\!\Psi_\Si[\phi_\Si^1,\dots,\phi_\Si^N,g_\Si]\overline{\Psi}_\Si[\phi_\Si'^1,\dots,\phi_\Si^N,g_\Si]\nn\\
&&\times\left(\prod_{n\neq1}\Dc\phi_\Si^n\right)\Dc g_\Si.
\eea
We insert $\Psib_\Si$ and $\overline{\Psib}_\Si$ from eq. \eqref{Psifshort}:
\bea\label{rhof}
\rho_\Si[\phi_\Si,\phi_\Si']&=&\int\exp\left(iNS_G[g]-iNS_G[g']\right)\nn\\
&&\times\Psi_\Si[\phi_\Si;g]\overline{\Psi}_\Si[\phi_\Si';g']~\langle\Psi_\Si;g'\ket{\Psi_\Si;g}^{N-1}\nn\\
&&\times\Psi^G_{\Sii}[g\Sii]\overline{\Psi}^G_{\Sii}[g'\Sii]\Dc g\Dc g',
\eea
where $\ket{\Psi_\Si;g}$ stands for the statevector of the wavefunctional $\Psi_\Si[\phi_\Si;g]$ \eqref{Psifsingle}. 
Because of the trace over the gravity subspace, 
it is understood that this time $g$ and $g'$ have the same final boundary values $g_\Si=g'_\Si$ and the path
integrals over $g,g'$ will extend for the final boundary $\Si$. 

Now we turn on the limit $\Nlim$. The factor $\langle\Psi_\Si;g'\ket{\Psi_\Si;g}^{N-1}$ vanishes if $g\not\equiv g'$.
Two avoid such degeneracy we assume that $g'-g=\delta g$ is a finite small function, then we take the limits $\Nlim$ and 
$\delta g\rightarrow0$ in this order. Using the leading order Taylor expansion 
$S_M[\Phi,g+\delta g]=-\tfrac12\int T^{ab}\delta g_{ab}\sqrt{\vert g\vert}dx$, we can derive the following relationship:  
\bea
&&\langle\Psi_\Si;g+dg\ket{\Psi_\Si;g}=\nn\\
&=&1+\frac{i}{2}\int\Sii^\Si\bra{\Psi\Sii}\To^{ab}_H(x)\ket{\Psi\Sii}\delta g_{ab}(x)\sqrt{\vert g\vert}dx.
\eea
The $(N-1)^{th}$ power in eq. \eqref{rhof} yields a phase factor diverging with $N$:
\bea
&&\langle\Psi_\Si;g+\delta g\ket{\Psi_\Si;g}^{N-1}=\\
&=&\exp\!\left(\! i\frac{N-1}{2}\!\!\int\!\bra{\Psi\Sii}\To^{ab}_H(x)\ket{\Psi\Sii}\delta g_{ab}(x)\sqrt{|g|}dx\!\right).\nn
\eea
Fortunately, there is another diverging phase on the r.h.s. of eq. \eqref{rhof}:
\bea
&&\exp\left(iNS_G[g]-iNS_G[g+\delta g]\right)=\nn\\
&=&\exp\left(-i\frac{N}{16\pi G}\int G^{ab}(x)\delta g_{ab}(x)\sqrt{|g|}dx\right).
\eea
For the two divergent phases to cancel each other out we must require that
\beq
G_{ab}(x)=8\pi G\bra{\Psi\Sii}\To_{ab}^H(x)\ket{\Psi\Sii},
\eeq
which is the semiclassical Einstein eq. \eqref{sclEE} in the Heisenberg picture.

Now we set $\delta g\equiv0$. The double path integral \eqref{rhof} reduces to a single one $\int\Dc g$.
It reduces further to the integral $\int\Dc g_\Si$ over the initial conditions since the rest of $g$ is determined by the 
semiclassical Einstein equation \eqref{sclEE}. The eq.  \eqref{rhof}  becomes  as simple as this:
\bea
\rho_\Si[\phi_\Si,\phi_\Si']
&=&\int\Psi_\Si[\phi_\Si;g]\overline{\Psi}_\Si[\phi_\Si';g]~\big{\vert}\Psi\Sii^G[g\Sii]\big{\vert}^2\Dc g\Sii\nn\\
\ro_\Si
&=&\int\ket{\Psi_\Si;g}\bra{\Psi_\Si;g}~\big{\vert}\Psi\Sii^G[g\Sii]\big{\vert}^2\Dc g\Sii,
\eea
where the lower equation re-writes the upper one into the Dirac formalism. 

According to this, the quantum state of the matter on hypersurface $\Si$ 
is the statistical mixture of pure states weighted by the probability distribution of the initial values
$g\Sii$ of the spacetime structure. If a single configuration $g\Sii$ is chosen then the quantum state
remains the pure state   $\Psi_\Si[\phi_\Si;g]$  \eqref{Psifsingle} which, as said there,  satisfies the 
Tomonaga--Schwinger equation \eqref{TSE}.  We already showed that the metric is determined
by the semiclassical Einstein equation \eqref{sclEE}.  This completes the proof that in the limit $\Nlim$
the emergent dynamics of any single cloneworld is semiclassical. 

\section{Correlated Worldlines}\label{III}
The original realization \cite{stamp2015rationale} of Stamp's concept
that infinite many (clone)fields are correlated by gravity 
has been changing over the years \cite{barvinsky2018structure}, the updated theory is
reviewed  in ref. \cite{wilson2022propagators}. 
There the generator functional for $N$ clones in the same quantized spacetime is defined by 
the following ring path integral:  
\beq\label{ZNJ}
Z_N[J]=\oint  e^{iNS_G[g]}\left(Z_1[g,J]\right)^N \Dc g,\\
\eeq
where 
\beq
Z_1[g,J]=\oint e^{iS_M[\phi,g]+i\int J\phi dx}\Dc\phi
\eeq
is the standard generator functional of a single field in fixed metric $g$. 
The  functional $Z_N[J]$ does not generate all correlations of the $N$ clones
$\phi^1,\phi^2,\dots,\phi^N$ on the same quantized metric $g$ 
but the correlations of the  collective variables $\sum_{n=1}^N\phi^n$. 
The functional $Z_N[J]$ generates what we
call the reduced dynamics of the summed field.
[We think that a plausible choice might be $Z_N[J/N]$, yielding the reduced dynamics of the \emph{average}
$(1/N)\sum_{n=1}^N\phi^n$ of the $N$ fields, avoiding 
divergences in the limit $\Nlim$.] But the CWL theory keeps on building. 
It constructs the above reduced dynamics of N-clones for $N=1,2,\dots,\infty$ 
and consider the uncorrelated composition of  all of them:
\beq\label{CWL}
Z[J]=\prod_{N=1}^\infty Z_N[J].
\eeq
This generator functional means a further reduction: it only generates the subdynamics
of the `particular collective variables', i.e.:  the sum of all fields, as observed in ref. \cite{barvinsky2018structure}.
%where  also mentioned that the unitarity in terms of the component fields is maintained.    

If we cut  the product at finite $N$ it contains $\nu_N=N(N-1)/2$ fields.
[We think again that plausible choice might be a straightforward rescaling of the current in $Z_N[J]$
in the above product yielding the reduced dynamics of the 
\emph{averages} of the $\nu_N$ fields to avoid divergencies in the limit $\Nlim$.]
CWL proposes the following rescaling of the genenator functional itself:  
\beq\label{CWLscaled}
Z_\mathrm{CWL}[J]=\prod_{N=1}^\infty \bigl(Z_N[J]\bigr)^{1/\nu_N}.
\eeq
In the ultimate form of CWL theory, this scaled generator constitutes the dynamics of 
the physical fields.  

However, the above rescaling is not standard in field theory. 
The unscaled generator  \eqref{CWL} described  (the limit $\Nlim$ of)  the  standard
reduced dynamics of the `particular collective variables',  legitimate in field theory at least formally. 
The rescaled generator is problematic.  %E.g., a constant pre-factor $1/\nu_N$ in front of the 
%connected correlations may not preserve the above mentioned unitarity since it resides on
%non-connected correlations as well. Furthermore, the rescaled generator
 It corresponds no more to the subdynamics of collective observables and it is unknown what the new observables could be.  
CWL theory \emph{postulates} that the rescaled generator generates the correlations
of  the physical fields.  Apparently, this interpretation is used  in applications as well 
\cite{wilson2024testing}. 

Just for comparison, let the generator functional representation  of the CCW theory (sec. \ref{II}) stand here:
\beq
Z_\mathrm{CCW}[J]=\lim_{\Nlim}\oint  e^{iNS_G[g]} Z_1[g,J]\left(Z_1[g,0]\right)^{N-1} \Dc g.
\eeq
This is equivalent with the reduced dynamics \eqref{rhof} upto the limit $\Nlim$  (the irrelavant $-1$ 
after $N$ is kept for full conformity). This generator functional formalism
offers an alternative way to see and prove how the infinite power of 
$Z_1[g,0]\equiv\langle\Psi_\Si;g'\ket{\Psi_\Si;g}$  under ring integral will make the metric $g$
classical.

The CCW theory is the $\Nlim$ limit of field theory of $N$ cloneworlds (N-CCW) 
of quantized matter in the same quantized spacetime. The physical world  is a standard reduction to one
of these replica worlds. 
The CWL theory is  the field theory of  the \emph{uncorrelated  composition} of all N-CCW from $N=1$ to $N=\infty$.
The physical world is a postulated dynamics, that does not follow from standard field theory.
CCW yield exact semiclassical gravity and it contains the remarkable self-attraction, 
best illustrated by the solitons of the the non-relativistic
\Schr--Newton equation \cite{diosi1984}. 
Self-attraction is a known mechanism of what is considered a key feature of `path bunching'  in CWL. 
Path bunching and semicalssical self-attraction are not exactly the same but very similar.  
That's not too surprising since CCW and CWL are based on related concepts and operate on  
similar mathematical structures.    

\section{Summary}
We showed that the  semiclassical gravity can be derived within standard quantum field theory
of infinite many copies of  the quantized matter, which we call (gravitationally)  correlated cloneworlds (CCW). 
Before his work with an infinite number of copies \cite{de2001schroedinger}, 
De Filippo discussed the Newton interaction with a single mirror of the quantized system, 
leading to entanglement between the physical world and its mirror \cite{de2001emergence}, see also
refs. \cite{maimone2023interaction,aguiar2024simple}.
The case of infinite number of clones is remarkable in that the entanglement between the clones disappears
asymptotically: we get the semiclassical Einstein (or the \Schr--Newton) equation for the single
physical world. The model requires cloneworlds which is a rather unnatural  technical assumption. 
Yet the merit of CCW stands: the semiclassical equations are not mere approximate equations
but exact consequences of standard quantum theory. 
This raises questions immediately since semiclassical gravity is long known to be 
inconsistent \cite{eppley1977}. A closer look shows that it is inconsistent with quantum measurements
and the statistical interpretation of the wavefunction  \cite{diosi2016nonlinear}. And, indeed, the CCW
theory can not accommodate measurements. The random measurement outcomes make the cloneworlds different
hence the post-measurement derivation of the semiclassical equations breaks down. 
Nevertheless, it is thought-provoking  whether non-selective measurements, 
or Everett's branchings instead \cite{everett1957relative}, 
could have some status in CCW --- and this way in semiclassical gravity.

\bibliography{diosi2024}{}

\begin{thebibliography}{17}%
\makeatletter
\providecommand \@ifxundefined [1]{%
 \@ifx{#1\undefined}
}%
\providecommand \@ifnum [1]{%
 \ifnum #1\expandafter \@firstoftwo
 \else \expandafter \@secondoftwo
 \fi
}%
\providecommand \@ifx [1]{%
 \ifx #1\expandafter \@firstoftwo
 \else \expandafter \@secondoftwo
 \fi
}%
\providecommand \natexlab [1]{#1}%
\providecommand \enquote  [1]{``#1''}%
\providecommand \bibnamefont  [1]{#1}%
\providecommand \bibfnamefont [1]{#1}%
\providecommand \citenamefont [1]{#1}%
\providecommand \href@noop [0]{\@secondoftwo}%
\providecommand \href [0]{\begingroup \@sanitize@url \@href}%
\providecommand \@href[1]{\@@startlink{#1}\@@href}%
\providecommand \@@href[1]{\endgroup#1\@@endlink}%
\providecommand \@sanitize@url [0]{\catcode `\\12\catcode `\$12\catcode
  `\&12\catcode `\#12\catcode `\^12\catcode `\_12\catcode `\%12\relax}%
\providecommand \@@startlink[1]{}%
\providecommand \@@endlink[0]{}%
\providecommand \url  [0]{\begingroup\@sanitize@url \@url }%
\providecommand \@url [1]{\endgroup\@href {#1}{\urlprefix }}%
\providecommand \urlprefix  [0]{URL }%
\providecommand \Eprint [0]{\href }%
\providecommand \doibase [0]{http://dx.doi.org/}%
\providecommand \selectlanguage [0]{\@gobble}%
\providecommand \bibinfo  [0]{\@secondoftwo}%
\providecommand \bibfield  [0]{\@secondoftwo}%
\providecommand \translation [1]{[#1]}%
\providecommand \BibitemOpen [0]{}%
\providecommand \bibitemStop [0]{}%
\providecommand \bibitemNoStop [0]{.\EOS\space}%
\providecommand \EOS [0]{\spacefactor3000\relax}%
\providecommand \BibitemShut  [1]{\csname bibitem#1\endcsname}%
\let\auto@bib@innerbib\@empty
%</preamble>
\bibitem [{\citenamefont {M{\o}ller~et al.}(1962)}]{moller1962}%
  \BibitemOpen
  \bibfield  {author} {\bibinfo {author} {\bibfnamefont {Christian}\
  \bibnamefont {M{\o}ller~et al.}},\ }\bibfield  {title} {\enquote {\bibinfo
  {title} {Les theories relativistes de la gravitation},}\ }\href@noop {}
  {\bibfield  {journal} {\bibinfo  {journal} {Colloques Internationaux CNRS}\
  }\textbf {\bibinfo {volume} {91}},\ \bibinfo {pages} {353} (\bibinfo {year}
  {1962})}\BibitemShut {NoStop}%
\bibitem [{\citenamefont {Rosenfeld}(1963)}]{rosenfeld1963}%
  \BibitemOpen
  \bibfield  {author} {\bibinfo {author} {\bibfnamefont {Leon}\ \bibnamefont
  {Rosenfeld}},\ }\bibfield  {title} {\enquote {\bibinfo {title} {On
  quantization of fields},}\ }\href@noop {} {\bibfield  {journal} {\bibinfo
  {journal} {Nuclear Physics}\ }\textbf {\bibinfo {volume} {40}},\ \bibinfo
  {pages} {1} (\bibinfo {year} {1963})}\BibitemShut {NoStop}%
\bibitem [{\citenamefont {Ruffini}\ and\ \citenamefont
  {Bonazzola}(1969)}]{ruffini1969systems}%
  \BibitemOpen
  \bibfield  {author} {\bibinfo {author} {\bibfnamefont {Remo}\ \bibnamefont
  {Ruffini}}\ and\ \bibinfo {author} {\bibfnamefont {Silvano}\ \bibnamefont
  {Bonazzola}},\ }\bibfield  {title} {\enquote {\bibinfo {title} {Systems of
  self-gravitating particles in general relativity and the concept of an
  equation of state},}\ }\href@noop {} {\bibfield  {journal} {\bibinfo
  {journal} {Physical Review}\ }\textbf {\bibinfo {volume} {187}},\ \bibinfo
  {pages} {1767} (\bibinfo {year} {1969})}\BibitemShut {NoStop}%
\bibitem [{\citenamefont {Di{\'o}si}(1984)}]{diosi1984}%
  \BibitemOpen
  \bibfield  {author} {\bibinfo {author} {\bibfnamefont {L.}~\bibnamefont
  {Di{\'o}si}},\ }\bibfield  {title} {\enquote {\bibinfo {title} {Gravitation
  and quantum-mechanical localization of macro-objects},}\ }\href {\doibase
  http://dx.doi.org/10.1016/0375-9601(84)90397-9} {\bibfield  {journal}
  {\bibinfo  {journal} {Physics Letters A}\ }\textbf {\bibinfo {volume}
  {105}},\ \bibinfo {pages} {199 -- 202} (\bibinfo {year} {1984})}\BibitemShut
  {NoStop}%
\bibitem [{\citenamefont {Penrose}(1996)}]{penrose1996}%
  \BibitemOpen
  \bibfield  {author} {\bibinfo {author} {\bibfnamefont {Roger}\ \bibnamefont
  {Penrose}},\ }\bibfield  {title} {\enquote {\bibinfo {title} {On gravity's
  role in quantum state reduction},}\ }\href@noop {} {\bibfield  {journal}
  {\bibinfo  {journal} {General Relativity and Gravitation}\ }\textbf {\bibinfo
  {volume} {28}},\ \bibinfo {pages} {581--600} (\bibinfo {year}
  {1996})}\BibitemShut {NoStop}%
\bibitem [{\citenamefont {Hartle}\ and\ \citenamefont
  {Horowitz}(1981)}]{hartle1981ground}%
  \BibitemOpen
  \bibfield  {author} {\bibinfo {author} {\bibfnamefont {James~B}\ \bibnamefont
  {Hartle}}\ and\ \bibinfo {author} {\bibfnamefont {Gary~T}\ \bibnamefont
  {Horowitz}},\ }\bibfield  {title} {\enquote {\bibinfo {title} {Ground-state
  expectation value of the metric in the 1/N or semiclassical approximation to
  quantum gravity},}\ }\href@noop {} {\bibfield  {journal} {\bibinfo  {journal}
  {Physical Review D}\ }\textbf {\bibinfo {volume} {24}},\ \bibinfo {pages}
  {257} (\bibinfo {year} {1981})}\BibitemShut {NoStop}%
\bibitem [{\citenamefont
  {De~Filippo}(2001{\natexlab{a}})}]{de2001schroedinger}%
  \BibitemOpen
  \bibfield  {author} {\bibinfo {author} {\bibfnamefont {Sergio}\ \bibnamefont
  {De~Filippo}},\ }\bibfield  {title} {\enquote {\bibinfo {title} {The
  Schroedinger-Newton model as N->infinity limit of a N color model},}\
  }\href@noop {} {\bibfield  {journal} {\bibinfo  {journal} {arXiv preprint
  gr-qc/0106057}\ } (\bibinfo {year} {2001}{\natexlab{a}})}\BibitemShut
  {NoStop}%
\bibitem [{\citenamefont {Stamp}(2015)}]{stamp2015rationale}%
  \BibitemOpen
  \bibfield  {author} {\bibinfo {author} {\bibfnamefont {PCE}\ \bibnamefont
  {Stamp}},\ }\bibfield  {title} {\enquote {\bibinfo {title} {Rationale for a
  correlated worldline theory of quantum gravity},}\ }\href@noop {} {\bibfield
  {journal} {\bibinfo  {journal} {New Journal of Physics}\ }\textbf {\bibinfo
  {volume} {17}},\ \bibinfo {pages} {065017} (\bibinfo {year}
  {2015})}\BibitemShut {NoStop}%
\bibitem [{\citenamefont {Barvinsky}\ \emph {et~al.}(2018)\citenamefont
  {Barvinsky}, \citenamefont {Carney},\ and\ \citenamefont
  {Stamp}}]{barvinsky2018structure}%
  \BibitemOpen
  \bibfield  {author} {\bibinfo {author} {\bibfnamefont {Andrei~O}\
  \bibnamefont {Barvinsky}}, \bibinfo {author} {\bibfnamefont {Daniel}\
  \bibnamefont {Carney}}, \ and\ \bibinfo {author} {\bibfnamefont {Philip~CE}\
  \bibnamefont {Stamp}},\ }\bibfield  {title} {\enquote {\bibinfo {title}
  {Structure of correlated worldline theories of quantum gravity},}\
  }\href@noop {} {\bibfield  {journal} {\bibinfo  {journal} {Physical Review
  D}\ }\textbf {\bibinfo {volume} {98}},\ \bibinfo {pages} {084052} (\bibinfo
  {year} {2018})}\BibitemShut {NoStop}%
\bibitem [{\citenamefont {Wilson-Gerow}\ and\ \citenamefont
  {Stamp}(2022)}]{wilson2022propagators}%
  \BibitemOpen
  \bibfield  {author} {\bibinfo {author} {\bibfnamefont {Jordan}\ \bibnamefont
  {Wilson-Gerow}}\ and\ \bibinfo {author} {\bibfnamefont {PCE}\ \bibnamefont
  {Stamp}},\ }\bibfield  {title} {\enquote {\bibinfo {title} {Propagators in
  the correlated worldline theory of quantum gravity},}\ }\href@noop {}
  {\bibfield  {journal} {\bibinfo  {journal} {Physical Review D}\ }\textbf
  {\bibinfo {volume} {105}},\ \bibinfo {pages} {084015} (\bibinfo {year}
  {2022})}\BibitemShut {NoStop}%
\bibitem [{\citenamefont {Wilson-Gerow}\ \emph {et~al.}(2024)\citenamefont
  {Wilson-Gerow}, \citenamefont {Chen},\ and\ \citenamefont
  {Stamp}}]{wilson2024testing}%
  \BibitemOpen
  \bibfield  {author} {\bibinfo {author} {\bibfnamefont {Jordan}\ \bibnamefont
  {Wilson-Gerow}}, \bibinfo {author} {\bibfnamefont {Yanbei}\ \bibnamefont
  {Chen}}, \ and\ \bibinfo {author} {\bibfnamefont {P.~C.~E.}\ \bibnamefont
  {Stamp}},\ }\bibfield  {title} {\enquote {\bibinfo {title} {Testing quantum
  gravity using pulsed optomechanical systems},}\ }\href {\doibase
  10.1103/PhysRevD.109.064078} {\bibfield  {journal} {\bibinfo  {journal}
  {Phys. Rev. D}\ }\textbf {\bibinfo {volume} {109}},\ \bibinfo {pages}
  {064078} (\bibinfo {year} {2024})}\BibitemShut {NoStop}%
\bibitem [{\citenamefont {De~Filippo}(2001{\natexlab{b}})}]{de2001emergence}%
  \BibitemOpen
  \bibfield  {author} {\bibinfo {author} {\bibfnamefont {Sergio}\ \bibnamefont
  {De~Filippo}},\ }\bibfield  {title} {\enquote {\bibinfo {title} {Emergence of
  classicality in non relativistic quantum mechanics through gravitational
  self-interaction},}\ }\href@noop {} {\bibfield  {journal} {\bibinfo
  {journal} {arXiv preprint quant-ph/0104052}\ } (\bibinfo {year}
  {2001}{\natexlab{b}})}\BibitemShut {NoStop}%
\bibitem [{\citenamefont {Maimone}\ \emph {et~al.}(2023)\citenamefont
  {Maimone}, \citenamefont {Naddeo},\ and\ \citenamefont
  {Scelza}}]{maimone2023interaction}%
  \BibitemOpen
  \bibfield  {author} {\bibinfo {author} {\bibfnamefont {Filippo}\ \bibnamefont
  {Maimone}}, \bibinfo {author} {\bibfnamefont {Adele}\ \bibnamefont {Naddeo}},
  \ and\ \bibinfo {author} {\bibfnamefont {Giovanni}\ \bibnamefont {Scelza}},\
  }\bibfield  {title} {\enquote {\bibinfo {title} {Interaction between everett
  worlds and fundamental decoherence in non-unitary Newtonian gravity},}\
  }\href@noop {} {\bibfield  {journal} {\bibinfo  {journal} {Universe}\
  }\textbf {\bibinfo {volume} {9}},\ \bibinfo {pages} {121} (\bibinfo {year}
  {2023})}\BibitemShut {NoStop}%
\bibitem [{\citenamefont {Aguiar}\ and\ \citenamefont
  {Matsas}(2024)}]{aguiar2024simple}%
  \BibitemOpen
  \bibfield  {author} {\bibinfo {author} {\bibfnamefont {Gabriel~HS}\
  \bibnamefont {Aguiar}}\ and\ \bibinfo {author} {\bibfnamefont {George~EA}\
  \bibnamefont {Matsas}},\ }\bibfield  {title} {\enquote {\bibinfo {title} {A
  simple gravitational self-decoherence model},}\ }\href@noop {} {\bibfield
  {journal} {\bibinfo  {journal} {arXiv preprint arXiv:2409.14155}\ } (\bibinfo
  {year} {2024})}\BibitemShut {NoStop}%
\bibitem [{\citenamefont {Eppley}\ and\ \citenamefont
  {Hannah}(1977)}]{eppley1977}%
  \BibitemOpen
  \bibfield  {author} {\bibinfo {author} {\bibfnamefont {Kenneth}\ \bibnamefont
  {Eppley}}\ and\ \bibinfo {author} {\bibfnamefont {Eric}\ \bibnamefont
  {Hannah}},\ }\bibfield  {title} {\enquote {\bibinfo {title} {The necessity of
  quantizing the gravitational field},}\ }\href@noop {} {\bibfield  {journal}
  {\bibinfo  {journal} {Foundations of Physics}\ }\textbf {\bibinfo {volume}
  {7}},\ \bibinfo {pages} {51--68} (\bibinfo {year} {1977})}\BibitemShut
  {NoStop}%
\bibitem [{\citenamefont {Di{\'o}si}(2016)}]{diosi2016nonlinear}%
  \BibitemOpen
  \bibfield  {author} {\bibinfo {author} {\bibfnamefont {Lajos}\ \bibnamefont
  {Di{\'o}si}},\ }\bibfield  {title} {\enquote {\bibinfo {title} {Nonlinear
  Schr{\"o}dinger equation in foundations: summary of 4 catches},}\ }\href@noop
  {} {\bibfield  {journal} {\bibinfo  {journal} {Journal of Physics: Conference
  Series}\ }\textbf {\bibinfo {volume} {701}},\ \bibinfo {pages} {012019}
  (\bibinfo {year} {2016})}\BibitemShut {NoStop}%
\bibitem [{\citenamefont {Everett~III}(1957)}]{everett1957relative}%
  \BibitemOpen
  \bibfield  {author} {\bibinfo {author} {\bibfnamefont {Hugh}\ \bibnamefont
  {Everett~III}},\ }\bibfield  {title} {\enquote {\bibinfo {title} {" `Relative
  state' formulation of quantum mechanics},}\ }\href@noop {} {\bibfield
  {journal} {\bibinfo  {journal} {Reviews of Modern Physics}\ }\textbf
  {\bibinfo {volume} {29}},\ \bibinfo {pages} {454} (\bibinfo {year}
  {1957})}\BibitemShut {NoStop}%
\end{thebibliography}

\end{document}